\documentclass[%
 reprint,
 amsmath,amssymb,
 aps, longbibliography,
showkeys,floatfix, pra, aps, amsmath, amssymb, superscriptaddress, twocolumn]{revtex4-2}
\usepackage{placeins}  
\usepackage{gensymb}
\usepackage{booktabs}
\usepackage{gensymb}
\usepackage{float}
\usepackage{graphicx}
\usepackage{dcolumn}
\usepackage{bm}
\usepackage{multirow}
\usepackage{color,outlines}
\usepackage{amsmath}
\usepackage{array}
\usepackage{physics}
\usepackage{hyperref}
\hypersetup{
    colorlinks=true,
    linkcolor=blue,
    citecolor=blue,
    urlcolor=blue
}

\begin{document}

\preprint{APS/123-QED}

\title{Precision Measurement of the Saturation Intensity in Rubidium at 420~nm}

\author{Shivam Sinha}
\thanks{These authors contributed equally to this work.}
\affiliation{Department of Physics, Indian Institute of Technology Tirupati, Yerpedu-517619, Andhra Pradesh, India.}

\author{Sumit Achar}
\thanks{These authors contributed equally to this work.}
\affiliation{Department of Physics, Indian Institute of Technology Tirupati, Yerpedu-517619, Andhra Pradesh, India.}

\author{Sankar Satheesh}
\affiliation{Department of Physics, Indian Institute of Technology Tirupati, Yerpedu-517619, Andhra Pradesh, India.}

\author{Arijit Sharma}
\email{arijit@iittp.ac.in}
\affiliation{Department of Physics, Indian Institute of Technology Tirupati, Yerpedu-517619, Andhra Pradesh, India.}
\affiliation{Center for Atomic, Molecular, and Optical Sciences and Technologies, Indian Institute of Technology Tirupati, Yerpedu-517619, Andhra Pradesh, India.}

\date{\today}

\begin{abstract}

The $5S_{1/2} \rightarrow 6P_{3/2}$ transition of rubidium (Rb) at 420~nm is a promising candidate for a portable warm-vapor all-optical atomic clock. Despite recent precision spectroscopy studies at 420 nm in Rb, an experimental determination of the saturation intensity of this transition has not yet been reported. The saturation intensity is a fundamental parameter that influences the  identification of a potential clock transition frequency in terms of optimizing various intensity-dependent parameters and connected systematics.  In this work, we report the first experimental measurement of the saturation intensity of the 420~nm transition in Rb, obtaining $(23.54 \pm 1.03)$~mW/cm$^2$ for the $^{87}$Rb $F=2\rightarrow F'=3$ transition and $(25.39 \pm 1.16)$~mW/cm$^2$ for the $^{85}$Rb $F=3\rightarrow F'=4$ transition, in excellent agreement with theoretical predictions. We further investigate the temperature dependence of the Doppler-free Lamb-dip amplitude and linewidth over 59.03$~\pm~0.37$ - 91.20$~\pm~0.90^\circ$C in a 100~mm commercial vapor cell, identifying around 82.02$~\pm~ 0.73^\circ$C as the optimal operating temperature, where the SNR of the Lamb-dip amplitude with temperature reaches a maximum, and the observed Lamb-dip linewidth exhibits a minimum. We also present precise measurements of the magnetic-dipole ($A$) and electric-quadrupole ($B$) hyperfine constants of the $6P_{3/2}$ state for both isotopes, with the measured values being consistent with previously reported values for the hyperfine constants.

\end{abstract}

\maketitle

\section{Introduction}

Alkali-metal atoms have played a central role in the development of precision spectroscopy \cite{macadam1992narrow, ovchinnikov2011accurate, 10.1063/1.4972567}, atomic frequency standards \cite{PhysRevLett.82.4619, 377790, bandi2022advanced, cui2024realization, 5006634, Muller:11, 84969}, quantum sensing \cite{bandi2023comprehensive,martin2018compact,obaze2025comprehensive}, and quantum information science \cite{hosseini2011high, Ryabtsev_2016, PhysRevA.77.032316, doi:10.1126/science.1110151}. Due to their relatively simple electronic structure and the availability of narrow-linewidth laser sources, Rb atoms are among the most extensively studied alkali-atomic systems for the realization of compact, field-deployable portable atomic clocks \cite{knappe2005, marlow9316270, Zhang:16}. Spectroscopic investigations of the D-line transitions $5S_{1/2}\rightarrow5P_{1/2}$ at 795 nm and $5S_{1/2}\rightarrow5P_{3/2}$ at 780 nm have led to numerous advances in laser stabilization \cite{xiang2009ultra}, atomic clocks \cite{guan2025780, gruet2017compact}, magnetometry \cite{budker2007optical}, and laser cooling \cite{cheng2009laser, van2014laser, zhang2024active, ludvigsen1994laser}. As a result, the spectroscopic properties of these transitions, including transition strengths, saturation intensities, hyperfine splittings, atomic number density,
and linewidth broadening mechanisms, are now well established \cite{steck2001rubidium,steck2008rubidium, achar2025direct}.\\

In contrast, transitions involving higher excited states have received comparatively lower attention despite their growing importance in portable atomic clock applications. Among these, the $5S_{1/2}\rightarrow6P_{3/2}$ transition at 420 nm is interesting because it provides a direct optical access to an excited state with a narrow linewidth (1.42~MHz \cite{J_Marek_1980, glaser2020absolute}), which opens up a viable pathway to realize an all-optical portable atomic clock \cite{10.1063/1.5006962, das2024direct, zhang2024power}. Additionally, the $6P_{3/2}$ state participates as one of the radiative decay channels for the two-photon excitation scheme via the ($5S_{1/2}\to5D_{5/2}$) transition \cite{noh2012transmittance}. Recent developments on two-photon optical atomic clocks use the 420 nm decay channel as the primary detection mechanism for the two-photon clock transition at 778.1 nm \cite{martin2018compact}. The 420 nm transition has also been employed in numerous studies involving excited-state spectroscopy \cite{zhang2014velocity}, laser cooling \cite{zhang2025diffuse}, blue fluorescence detection, frequency up-conversion processes \cite{vernier2010enhanced}, Rydberg-atom excitation pathways \cite{xu2025tunable}, precision measurements of atomic structure \cite{das2024direct,navarro2019doppler,glaser2020absolute,arimondo1977experimental,safronova2011critically}, and optical atomic clocks \cite{zhang2017420, zhang2024power}. Furthermore, the shorter wavelength and larger transition energy make this transition attractive for applications requiring enhanced spectroscopic resolution and state-selective detection \cite{glaser2020absolute, PhysRevA.92.042511}.\\

To optimize the use of the 420 nm transition as a viable optical clock excitation scheme, it is important to have an accurate understanding of its saturation intensities. It is a critical parameter that influences the  identification of a potential clock transition frequency in terms of optimizing various intensity-dependent parameters and connected systematics. Without an accurate knowledge of the saturation intensity for a given transition, the optimal operating point, the sensitivity limit, the systematic frequency correction, and the calibration of the absolute output cannot be determined precisely for any kind of optical frequency standard using the transition. Interestingly, once the atom is excited to the $6P_{3/2}$ state via the 420 nm transition, there are multiple decay channels available for de-excitation to the ground state $5S_{1/2}$. In fact, the branching ratio for the atom to decay along the $6P_{3/2}\to5S_{1/2}$ state is only around 23$\%$ \cite{PhysRevA.69.022509, noh2012transmittance}, which limits the SNR for all detection mechanisms based on this channel.

The saturation intensity is the optical intensity at which the excitation rate of an atomic transition becomes comparable to the spontaneous decay rate \cite{milonni2010laser}. It determines the optical power required to significantly modify the atomic population distribution and governs a wide range of nonlinear spectroscopic phenomena, including power broadening, optical pumping, and saturated absorption spectroscopy. The knowledge of the saturation intensity is essential for the design and optimization of laser-spectroscopy experiments, frequency references, and quantum sensors. For the Rb D-line transitions, saturation intensities are well known and widely used \cite{haupl2025modelling, Akulshin1990, steck2001rubidium}, the corresponding values for the $5S_{1/2}\rightarrow6P_{3/2}$ transition at 420~nm remain comparatively unexplored. Notably, the reported saturation intensity of the 420 nm transition in Rb remains ambiguous in the literature, with previously reported values spanning more than an order of magnitude \cite{10.1063/1.5006962,10.1088/1674-1056/ae194c, vernier2010enhanced}. The presence of multiple decay channels from the $6P_{3/2}$ state further complicates the determination of an effective transition strength and therefore an accurate determination of the corresponding saturation intensity.\\


In this work, we investigate the 420~nm $5S_{1/2}\rightarrow6P_{3/2}$ transition of Rb using saturated absorption spectroscopy (SAS) \cite{glaser2020absolute,lee2009reversed,jacques2009nonlinear,das2024direct}. A theoretical calculation of the saturation intensity for the 420~nm transition in Rb has been performed using a first-principles approach. The predicted saturation intensities are then compared with the experimental values obtained from power-broadening measurements of the Lamb dip resonances using SAS for both $^{85}$Rb and $^{87}$Rb isotopes. The zero-intensity extrapolation gives $\Gamma_0$ (the zero-power linewidth), and $I_{\text{sat}}$ is extracted by fitting the power broadening equation to $\Gamma_{h}$ versus power plot, and we are able to achieve less than 1$\%$ accuracy with careful lineshape analysis. To the best of our knowledge, this represents the first experimental measurement of the saturation intensity parameters for the 420~nm transitions in Rb.\\

In addition to the saturation-intensity measurements, we investigated the thermal dependence of the Lamb dip amplitude and linewidth using the Doppler-free saturated absorption signal at 420~nm. The SNR and resonance line width of the Lamb dips are studied over a broad temperature range, revealing an optimum operating temperature near 82.02$~\pm~ 0.73^\circ$C. We also present precise measurements of the hyperfine structure constants of the $6P_{3/2}$ state, which are extracted from the measured SAS spectra and compared with previously reported values.\\

The results presented here provide a quantitative analysis of the saturation behavior of the 420~nm rubidium transition and establish experimentally validated saturation-intensity values for the strongest hyperfine transitions of both naturally occurring rubidium isotopes. These measurements are expected to be useful for future investigations involving blue-transition spectroscopy, precision metrology, frequency stabilization, and quantum technologies based on higher-excited-state alkali atoms.\\

The manuscript is organized as follows. Section~\ref{sec:theory} presents 
the theoretical framework for determining the saturation intensity parameters over various hyperfine transitions at 420~nm for the $6P_{3/2}$ manifold, including the branching ratio corrections. Section~\ref{sec:experiment} describes the experimental 
setup, including the laser system, beam geometry, vapor cell configuration, temperature stabilization, and frequency calibration procedure. Section~\ref{sec:results} presents the experimental results: the measurement of saturation intensities for both isotopes, the temperature dependence of the Lamb-dip amplitude and linewidth, and the measurement of hyperfine constants. Section~\ref{sec:conclusion} summarizes the main findings and discusses 
their implications for future applications.

\section{Theoretical background \label{sec:theory}}
In this section, the two theoretical quantities measured in this work are discussed: 
the saturation intensity and the hyperfine structure constants of the 
$6P_{3/2}$ state. We first derive the saturation intensity of a hyperfine 
transition starting from the two-level atom result and building up the dipole matrix element via the Wigner-Eckart theorem \cite{brink1994angular, Sakurai_Napolitano_2017}, 
and applying a branching-ratio correction specific to the open decay 
topology of the 420~nm transition. We then outline the hyperfine 
interaction Hamiltonian used to relate the measured Lamb-dip frequency 
intervals to the magnetic-dipole ($A$) and electric-quadrupole ($B$) 
coupling constants.
\subsection{Saturation Intensity}
The saturation intensity $I_{\text{sat}}$, is a fundamental parameter in light-matter interactions. It is formally defined as the optical intensity at which the transition probability becomes comparable to the relaxation rate of the atomic system, such that the population in the excited state reaches exactly half of its maximum possible steady-state value \cite{metcalf1999laser}.\\ 

By comparing the total photon scattering rate with the steady-state population of the excited state in an idealized two-level atomic system, the relationship between the incident laser intensity $I$ and $I_{\text{sat}}$ is given by \cite{steck2008rubidium}:
\begin{equation}
    \frac{I}{I_{\text{sat}}} = 2 \left( \frac{\Omega}{\Gamma} \right)^2,
    \label{eq:isat_rabi}
\end{equation}
where $\Omega$ represents the Rabi frequency and $\Gamma = 1/\tau$ is the natural decay rate (or inverse radiative lifetime $\tau$) of the excited state. \\

Given that the intensity of an electromagnetic plane wave is related to its electric field amplitude $E_0$ by $I = \frac{1}{2}c\epsilon_0 E_0^2$, using Eq.~(\ref{eq:isat_rabi}), the saturation intensity can be expressed as:
\begin{equation}
    I_{\text{sat}} = \frac{c \epsilon_0 \Gamma^2 \hbar^2}{4 |\hat{\epsilon} \cdot \mathbf{d}|^2},
    \label{eq:isat_dipole}
\end{equation}
where $c$ is the speed of light in vacuum, $\epsilon_0$ is the vacuum permittivity, $\hat{\epsilon}$ is the unit polarization vector of the incident light field, and $\mathbf{d}$ is the atomic dipole moment operator.\\

To evaluate $I_{\text{sat}}$, the dipole moment in Eq.~(\ref{eq:isat_dipole}) must be replaced by the specific transition matrix elements linking individual hyperfine magnetic sub-levels $\ket{F, m_F}$ and $\ket{F', m_F'}$. Using the Wigner-Eckart theorem \cite{brink1994angular, Sakurai_Napolitano_2017}, the matrix element for a light field with polarization $q$ (where $q = 0, \pm 1$) can be factored into a reduced matrix element and a Wigner 3-j symbol \cite{steck2001rubidium},

\begin{equation}
    \begin{split}
            \bra{F, m_F} e r_q \ket{F', m_F'} &= \bra{F} \norm{er} \ket{F'} (-1)^{F' - 1 + m_F} \sqrt{2F+1}\\
            & \times \begin{pmatrix} F' & 1 & F \\ m_F' & q & -m_F \end{pmatrix}.
            \label{eq:3j}
    \end{split}
\end{equation}

The dependency on the total nuclear spin $I$ and the total electronic angular momentum $J$ hidden within the hyperfine reduced matrix element $\bra{F} \norm{er} \ket{F'} \equiv \bra{J, I, F} \norm{er} \ket{J', I', F'}$ appearing in Eq.~(\ref{eq:3j}) can be further isolated by utilizing the Wigner 6-j symbols:
\begin{equation}
    \begin{split}
            \bra{J, I, F} \norm{er} \ket{J', I', F'} &= \bra{J} \norm{er} \ket{J'} (-1)^{F' + J + 1 + I} \\
            &\times \sqrt{(2F'+1)(2J+1)} \begin{Bmatrix} J & J' & 1 \\ F' & F & I \end{Bmatrix}.
            \label{eq:6j_hyperfine}
    \end{split}
\end{equation}

Similarly, the reduced fine structure matrix element $\bra{J} \norm{er} \ket{J'} \equiv \bra{L, S, J} \norm{er} \ket{L', S', J'}$ appearing in Eq.~(\ref{eq:6j_hyperfine}) can be broken down to isolate the fine structure spin $S$ and the orbital angular momentum $L$:
\begin{equation}
\begin{split}
        \bra{L, S, J} \norm{er} \ket{L', S', J'} &= \bra{L} \norm{er} \ket{L'} (-1)^{J' + L + 1 + S}\\
        & \times  \sqrt{(2J'+1)(2L+1)} \begin{Bmatrix} L & L' & 1 \\ J' & J & S \end{Bmatrix}.
        \label{eq:6j_fine}
\end{split}
\end{equation}

By Combining Eqs.~(\ref{eq:3j})-(\ref{eq:6j_fine}), The overall effective square of the dipole coupling strength $|d_{\text{eff}}|^2$ for a specific transition between states is structured as:
\begin{equation}
    \begin{split}
            |d_{\text{eff}}|^2 &= (2F+1)(2F'+1)(2J+1)(2J'+1)(2L+1) \\
    &\times \begin{pmatrix} F' & 1 & F \\ m_F' & q & -m_F \end{pmatrix}^2 \begin{Bmatrix} J & J' & 1 \\ F & F' & I \end{Bmatrix}^2 \begin{Bmatrix} L & L' & 1 \\ J & J' & S \end{Bmatrix}^2\\
    &\times \bra{L}\norm{er} \ket{L'}^2,
    \label{eq:deff}
    \end{split}
\end{equation}
where $\bra{L} \norm{er} \ket{L'}$ is the reduced dipole matrix element between the ground state $S$ ($L_g = 0$) and the excited state $P$ ($L_e = 1$). For an electric-dipole transition between an excited
state $\ket{i}$ and a lower state $\ket{f}$, the reduced matrix element
is related to the Einstein spontaneous-emission coefficient $A_{if}$ by \cite{siddons2008absolute,loudon2000quantum},

\begin{equation}
\bra{L_g=0}\norm{er}\ket{L_e=1}
=
\sqrt{3}
\sqrt{
\frac{
3\epsilon_0\hbar A_{if}\lambda^3
}
{
8\pi^2
}
},
\label{eq:reduced_matrix}
\end{equation}

where $\lambda$ is the transition wavelength and $A_{if}$ denotes the
Einstein's spontaneous-emission coefficient, i.e., the probability per unit
time for spontaneous decay from the excited state $\ket{i}$ to the lower
state $\ket{f}$. For a closed two-level transition, the excited state
decays through a single radiative channel and therefore $A_{if}=A_{\rm tot}=\Gamma,$ where $A_{\rm tot}$ (or equivalently $\Gamma$) is the total spontaneous decay rate or natural linewidth of the excited state.\\

Eq.~(\ref{eq:deff}), together with Eq.~(\ref{eq:reduced_matrix}), 
fully specifies $|d_{\text{eff}}|^2$ for a transition between a particular Zeeman sub-levels $m_F \to m_F'$. To obtain the saturation intensity 
that is independent of the specific sub-level and polarization chosen, 
the $m_F$-resolved expression for $|d_{\rm eff}|^2$ must be summed over the polarization components $q$, as described below.\\

In the absence of an external magnetic field, the individual Zeeman sub-levels remain degenerate, and the light field interacts equally with all magnetic field sub-levels. Due to the inherent symmetries of the dipole operator, all excited-state sub-levels decay at the identical overall rate $\Gamma$. The population leaving the excited state branches out into accessible ground-state sub-levels according to the line strength factor $S_{FF'}$ \cite{steck2001rubidium}, defined via the summation over all possible polarizations $q=0,\pm~1$ of the squared Wigner 3-j symbol.\\


Following the isotropic averaging formalism \cite{steck2001rubidium,steck2008rubidium}, the effective dipole moment for the hyperfine transition $F \rightarrow F'$ transitions is given by,
\begin{equation}
    |d_{\text{iso,eff}}(F \rightarrow F')|^2 = \frac{1}{3} S_{FF'} \left| \bra{J} \norm{er} \ket{J'} \right|^2.
    \label{eq:diso_compact}
\end{equation}
Eq.~(\ref{eq:diso_compact}) can be written in its expanded form relative to the orbital angular momentum basis, which gives,
\begin{equation}
\begin{split}
        |d_{\text{iso,eff}}(F \rightarrow F')|^2 &= \frac{1}{3} S_{FF'} (2J'+1)(2L+1) \begin{Bmatrix} J & J' & 1 \\ F & F' & I \end{Bmatrix}^2 \\
        & \times \bra{L_g=0} \norm{er} \ket{L_e=1}^2.
    \label{eq:diso_expanded}
\end{split}
\end{equation}
Substituting the isotropic effective dipole coupling (from Eq.~(\ref{eq:diso_expanded})) back into the saturation intensity equation Eq.~(\ref{eq:isat_dipole}) yields the closed-form expression for the hyperfine transition's saturation intensity,
\begin{equation}
    I_{\text{sat}} = \frac{c \epsilon_0 \Gamma^2 \hbar^2}{4 |d_{\text{iso,eff}}(F \rightarrow F')|^2}.
    \label{eq:isat_final}
\end{equation}

The well-studied Rb $D_2$ line ($5S_{1/2} \rightarrow 5P_{3/2}$) at 780 nm, the excited atoms decay almost exclusively back to the initial ground states. This yields highly optimized coupling and leads to low saturation intensities. Using Eq.~(\ref{eq:isat_final}), we obtain the $I_{\text{sat}} = 3.57\text{ mW/cm}^2$ for $^{87}\text{Rb}~(F=2 \rightarrow F'=3$ transition) \cite{steck2001rubidium} and $3.89\text{ mW/cm}^2$ for $^{85}\text{Rb}~(F=3 \rightarrow F'=4$ transition) \cite{steck2008rubidium}.\\

\begin{figure}[h!]
    \centering
    \includegraphics[width=0.8\linewidth]{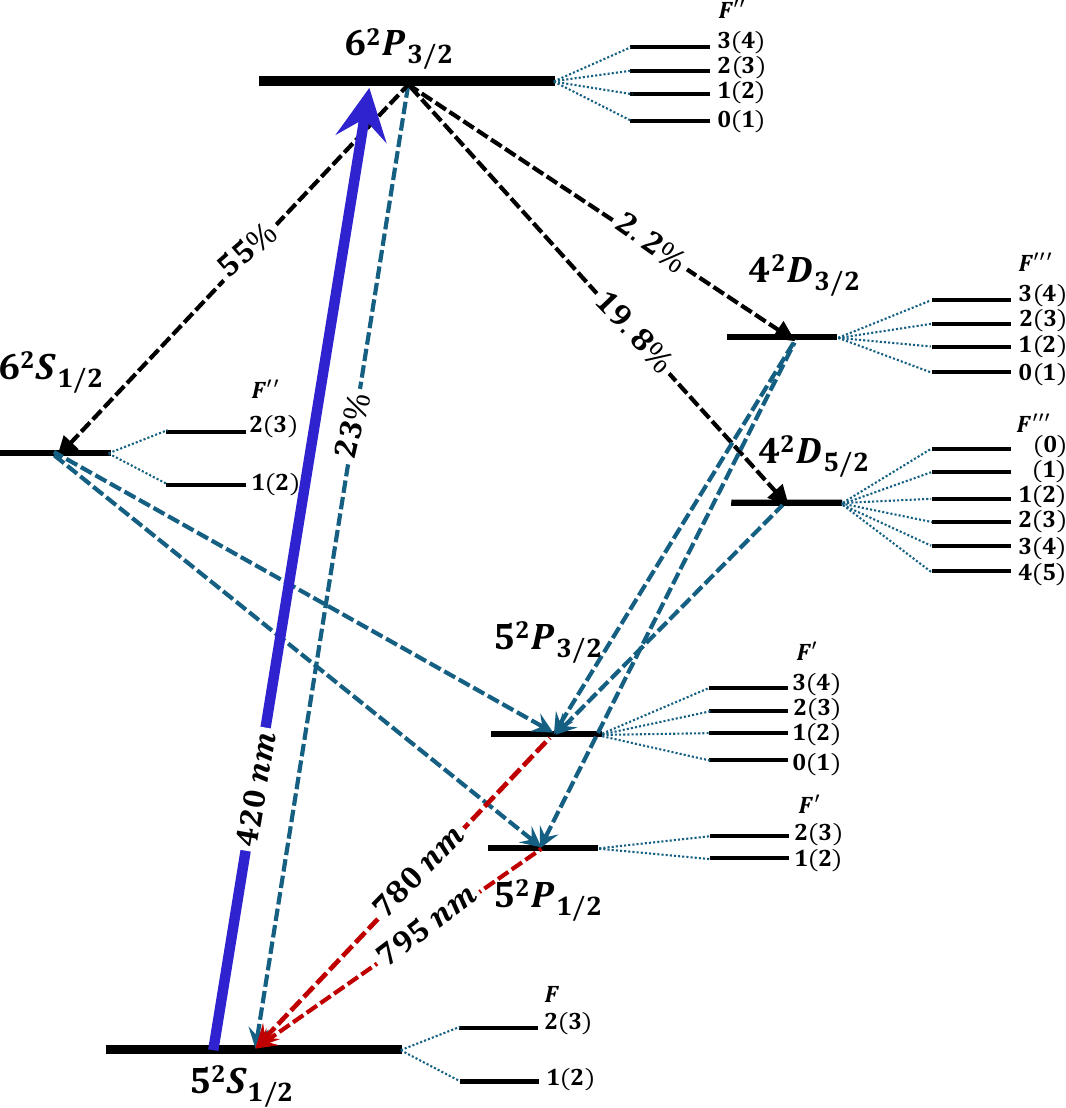}
    \caption{Relevant energy-level structure of the rubidium $5^{2}S_{1/2}\rightarrow6^{2}P_{3/2}$ transition at 420~nm. Hyperfine levels labeled without parentheses represent $^{87}$Rb, while those in parentheses represent $^{85}$Rb. The hyperfine transitions investigated in this work are indicated in a solid blue color. The spontaneous decay pathways from the $6P_{3/2}$ state and their associated branching ratios are also shown.}
    \label{fig:energy_levels}
\end{figure}

However, for the $5S_{1/2}\rightarrow6P_{3/2}$ transition at 420~nm, the $6P_{3/2}$ state decays through several radiative channels involving the $6S_{1/2}$, $4D_{3/2}$ and $4D_{5/2}$ manifolds before ultimately returning to the
ground state $5S_{1/2}$, as illustrated in Fig.~\ref{fig:energy_levels}. Consequently, the total spontaneous decay rate is given by
\begin{equation}
A_{\rm tot}=\sum_k A_{ik},
\end{equation}
where the summation extends over all allowed spontaneous emission
channels. Only approximately 23\% of the total spontaneous decay from the
$6P_{3/2}$ state proceeds directly through the driven
$6P_{3/2}\rightarrow5S_{1/2}$ transition
\cite{noh2012transmittance}. The Einstein coefficient relevant for the
420~nm transition is therefore the partial spontaneous-emission rate

\begin{equation}
A_{420}=A(6P_{3/2}\rightarrow5S_{1/2})=\beta A_{\rm tot},
\end{equation}

where $\beta\simeq0.23$ is the branching ratio associated with the
420~nm decay channel.

Accordingly, the reduced dipole matrix element in
Eq.~(\ref{eq:reduced_matrix}) is evaluated using the partial Einstein
coefficient $A_{420}$ rather than the total decay rate. It modifies the effective transition coupling strength by scaling down the coherent dipole cross-section:
\begin{equation}
    |d_{\text{eff}}|^2 \longrightarrow 0.23 \times |d_{\text{eff}}|^2.
\end{equation}


Since the saturation intensity varies inversely with the square of the effective
dipole matrix element, the reduced transition probability associated
with the open decay channels leads to a substantially larger saturation
intensity than that of an equivalent closed transition.\\

Using the above formalism, the theoretical saturation intensities for
the two strongest hyperfine transitions at 420~nm are calculated to be,
\begin{equation}
    I_{\text{sat, theo}}(^{87}\text{Rb}, F=2 \rightarrow F'=3) = 23.45\text{ mW/cm}^2,
\end{equation}
\begin{equation}
    I_{\text{sat, theo}}(^{85}\text{Rb}, F=3 \rightarrow F'=4) = 25.54\text{ mW/cm}^2.
\end{equation}
These values serve as the primary baseline benchmarks validated across our multi-temperature experimental profiles within the vapor cell environment.

\subsection{Hyperfine Structure and Energy Levels}
The hyperfine structure arises from the interaction between the atomic total angular momentum $\mathbf{J}$ and the nuclear angular momentum $\mathbf{I}$ \cite{woodgate1969elementary, foot2005atomic}. This interaction is significantly weaker than the fine-structure coupling due to the relatively small magnitude of the nuclear magnetic moment $\boldsymbol{\mu}_I$, defined as:
\begin{equation}
    \boldsymbol{\mu}_I = - \mu_N g_I \frac{\mathbf{I}}{\hbar}
\end{equation}
where $\mu_N$ is the nuclear magneton and $g_I$ is the nuclear $g$-factor, and $h$ is the Planck constant. At the nucleus, the valence electron generates an effective magnetic field $\mathbf{B}$, which is proportional to its total angular momentum $\mathbf{J}$:
\begin{equation}
    \mathbf{B} = -b \mathbf{J},
\end{equation}
where $b$ is a positive constant determined by the electronic wave function at the nucleus. The hyperfine interaction Hamiltonian is defined by the energy of the nuclear magnetic moment in the electronic magnetic field \cite{arimondo1977experimental},
\begin{equation}
    H_{\text{hfs}} = -\boldsymbol{\mu}_I \cdot \mathbf{B} = A \frac{\mathbf{I} \cdot \mathbf{J}}{\hbar^2},
\end{equation}
where $A$ is the magnetic dipole hyperfine coupling constant. Including higher-order effects, the Hamiltonian is expanded into a multipole series. Beyond the magnetic dipole term (multipole rank $k=1$), the dominant contribution for states with $J > 1/2$ is the electric quadrupole interaction (multipole rank $k=2$), characterized by the constant $B$. The full Hamiltonian is:
\begin{equation}
\begin{split}
H_{\mathrm{hfs}} =\;&
A h \mathbf{I}\cdot\mathbf{J} \\
&+ B h\,
\frac{
3(\mathbf{I}\cdot\mathbf{J})^2
+\frac{3}{2}(\mathbf{I}\cdot\mathbf{J})
-I(I+1)J(J+1)
}{
2I(2I-1)J(2J-1)
}.
\end{split}
\end{equation}

Defining the quantum number $K = F(F+1) - I(I+1) - J(J+1)$, where $\mathbf{F} = \mathbf{I} + \mathbf{J}$ is the total atomic angular momentum, the energy shift $\Delta E_{\text{hfs}}$ for a state $|J, I, F \rangle$ is given by \cite{arimondo1977experimental, steck2001rubidium}:
\begin{equation}
    \Delta E_{\text{hfs}} = \frac{1}{2}h A K + B h \, \frac{\frac{3}{2}K(K+1) - 2I(I+1)J(J+1)}{2I(2I-1)2J(2J-1)}.
\end{equation}
The magnetic dipole term $A$ arises from the spin-spin coupling between the nuclear magnetic moment and the electronic magnetic field, reflecting the magnetic interaction of the atom. The quadrupole term $B$ originates from the nuclear electric quadrupole moment, which characterizes the deviation of the nuclear charge distribution from spherical symmetry. In rubidium, this term is essential for accurately mapping the $6P_{3/2}$ energy levels, as it accounts for subtle energy shifts arising from non-spherical nuclear charge distributions. By experimentally resolving the hyperfine frequency intervals and fitting them to this Hamiltonian, we extract the precise values of $A$ and $B$ for the $6P_{3/2}$ transition.

\section{Experimental section \label{sec:experiment}}
\begin{figure}[h!]
    \centering
\includegraphics[width=\linewidth]{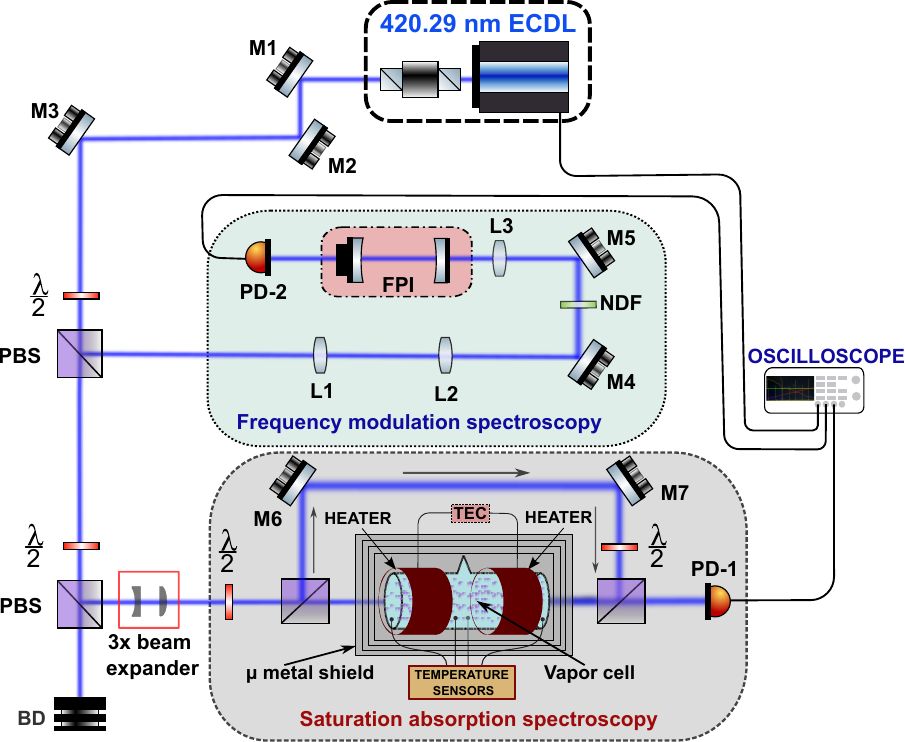}
    \caption{Experimental schematic of the saturation absorption spectroscopy of a 100 mm long Rb vapor cell on the $5S_{1/2}\rightarrow6P_{3/2}$ transition using a 420.29 nm laser. The vapor cell is placed inside a 4-layer $\mu$-metal shield, and the spectroscopy beam is expanded by a factor of three before entering the cell. The transmitted signal through the vapor cell is detected using a photodetector (PD-1), while the reference transmission signal from the Fabry–Pérot interferometer (FPI) is monitored by photodetector (PD-2). The outputs of both photodetectors are simultaneously recorded with a digital storage oscilloscope triggered by the ECDL piezoelectric transducer scan signal. The vapor cell temperature is stabilized using a temperature controller and resistive foil heaters, which are monitored by four NTC  temperature sensors. M1–M7, mirrors; L1–L3, lenses; PBS, polarizing beam splitter; $\lambda/2$, half-wave plate; NDF, neutral-density filter; BD, beam dump.}
    \label{fig:experimental_setup}
\end{figure}

The experimental setup used to perform SAS \cite{preston1996doppler} of the rubidium $5S_{1/2}\rightarrow6P_{3/2}$ transition at 420~nm is shown in Fig.~\ref{fig:experimental_setup}. A narrow-linewidth external-cavity diode laser (ECDL) \cite{macadam1992narrow, ricci1995compact} of 420.29 nm was used to interrogate the transition. The laser frequency was scanned across the hyperfine resonances of both naturally occurring rubidium isotopes, $^{85}$Rb and $^{87}$Rb.\\

The laser output was characterized using a calibrated optical power meter, and the spatial beam profile was monitored using a CCD camera to ensure near-TEM$_{00}$ mode operation. The beam diameters ($1/e^2$) were approximately $2.985 \pm 0.05$ mm and $1.955 \pm 0.04$ mm along the major and minor axes of the ellipse, respectively. The optical intensity was calculated from the measured power using the effective beam area, $A=\frac{\pi w_x w_y}{4}$, where $w_x$ and $w_y$ denote the $1/e^2$ beam diameters along the major and minor axes of the elliptical beam.\\

The laser beam was directed through a half-wave plate ($\lambda/2$) and a polarizing beam splitter (PBS), which were used to control the optical power and generate the counter-propagating pump and probe beams required for SAS. The probe power was maintained at approximately $\sim$50~$\mu$W and the pump power was varied from $\sim$300~$\mu$W to $\sim$3~mW. The probe beam was kept sufficiently weak to minimize probe-induced saturation and ensure operation in the weak-probe regime.\\

To assess the robustness of the measurement, the saturation intensity was determined using both beam configurations and for several vapor-cell temperatures (49.35$~\pm~ 0.25^\circ$C, 59.03$~\pm~ 0.37^\circ$C, 72.90$~\pm~ 0.57^\circ$C, and 82.02$~\pm~ 0.73^\circ$C). Since the saturation intensity is an intrinsic property of the atomic transition, it is expected to be independent of beam size and operating temperature. Repeating the measurements under different experimental conditions provided an important consistency check.\\

The vapor cell was mounted in a custom-designed heating assembly and enclosed within a four-layer $\mu$-metal magnetic shield to suppress the influence of the Earth's ambient magnetic field. The heating assembly was regulated using a Thorlabs TC300B temperature controller. To minimize rubidium condensation on the optical windows, the heating elements were positioned near the corners of the vapor cell, leaving the central region free of direct heating. Since this configuration can introduce temperature gradients across the vapor cell, four pre-calibrated negative temperature coefficient (NTC) sensors were attached at different locations: two at the corners and two near the center of the vapor cell. The sensor outputs were continuously monitored using an Arduino Uno microcontroller, and the average of the four readings was taken as the effective vapor-cell temperature throughout the experiment. The temperature controller was set to 50$^\circ$C, 60$^\circ$C, 75$^\circ$C, and 85$^\circ$C the corresponding average measured temperatures were 49.35$~\pm~ 0.25^\circ$C, 59.03$~\pm~ 0.37^\circ$C, 72.90$~\pm~ 0.57^\circ$C, and 82.02$~\pm~ 0.73^\circ$C respectively, for the measurement of the saturation intensity.\\

The SAS signal was detected using an unbiased, amplified photodetector (Thorlabs PDA100A2) and recorded on one channel of a 4-channel digital storage oscilloscope (Tektronix MSO44B). For the frequency calibration, a fraction of the laser beam was directed through a lens system (L1 and L2) and mode-matched into a Fabry-Pérot interferometer (Thorlabs SA30-47 with FSR = $1.5075$ GHz) \cite{bjorklund1980frequency, bjorklund1983frequency} using lens L3. The transmitted signal from the interferometer was captured by another photodetector (PD-2) and recorded on a separate channel of the oscilloscope. The SAS signal and the interferometer transmission signal were acquired simultaneously within the same acquisition window, ensuring synchronized data acquisition for accurate analysis. \\

The non-linearities in the piezoelectric laser frequency scan were corrected by mapping the FPI transmission peaks to their known frequency intervals \cite{Pizzey_2022, achar2025direct}. The power scan non-linearities, inherent in the diode laser current-tuning, were removed by fitting the off-resonant regions of the transmission spectra with a higher-order polynomial; this polynomial characterization was then used to normalize the atomic absorption and saturated absorption signals, to ensure that the measured Lamb-dip amplitudes and widths were strictly functions of the atomic transition dynamics rather than artifacts of the experimental scan.\\
\begin{figure}[h!]
    \centering
    \includegraphics[width=1\linewidth]{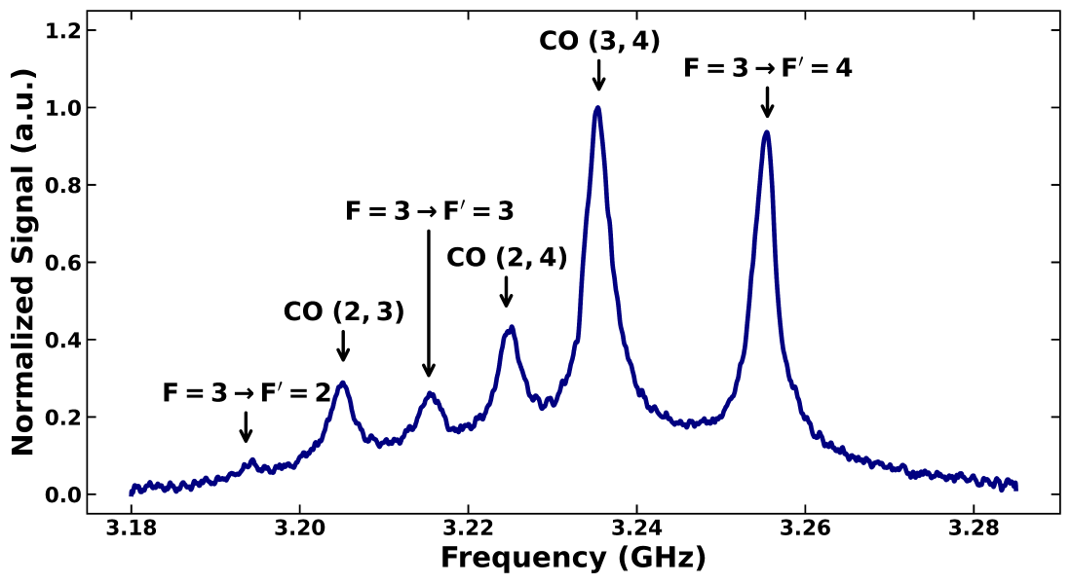}
    \caption{Doppler-free saturated absorption spectrum of the $^{85}$Rb $5S_{1/2}(F=3)\rightarrow6P_{3/2}(F' =2,3,4)$ transitions at 82.02$~\pm~ 0.73^\circ$C, recorded with a probe power of $\sim$50~$\mu$W and a pump power of $\sim$1100~$\mu$W. The observed Lamb dips and crossover resonances (CO) are indicated.}
    \label{fig:sas_rb_85}
\end{figure}

To extract the hyperfine magnetic dipole ($A$) and electric quadrupole ($B$) coefficients of the $6P_{3/2}$ state, we conducted high-resolution saturated absorption measurements at a stabilized temperature of 82.02$~\pm~0.73^{\circ}$C. Using an optimized configuration with a probe power of $\sim$50~$\mu$W and a pump power of $\sim$1100~$\mu$W, the measurement was repeated four times, with 50 independent datasets acquired in each repetition, to ensure high statistical confidence.\\

The Doppler-free spectral resolution achieved at this power regime allowed for the distinct identification of the hyperfine transition manifolds. Fig. \ref{fig:sas_rb_85} illustrates a representative saturated absorption spectrum for the $^{85}$Rb ($F=3 \rightarrow F'=4$) transition, highlighting the well-resolved Lamb-dip features. By mapping these spectral peaks to the absolute frequency scale provided by the synchronized Fabry-Pérot interferometer transmission, we precisely determined the hyperfine intervals \cite{arimondo1977experimental}. These measured intervals served as the input for a least-squares fit to the hyperfine Hamiltonian, allowing for the extraction of the $A$ and $B$ coefficients for the $6P_{3/2}$ excited state.

\section{Results}\label{sec:results}

We first focus on the measurement of $I_{\rm sat}$ using 
the power-broadening of the Lamb-dip linewidth, followed by a characterization 
of how the cell temperature governs the trade-off between signal amplitude and 
linewidth. We then turn to the hyperfine structure of the $6P_{3/2}$ state, 
where the well-resolved Lamb dips serve as precise frequency markers.


\subsection{Measurement of Saturation Intensity}
The saturation intensity of the $5S_{1/2} \rightarrow 6P_{3/2}$ transition 
at 420.29~nm was measured for both $^{85}$Rb and $^{87}$Rb using the 
power-broadening of the Lamb-dip linewidth. The measurement procedure 
consists of two sequential steps.\\

In the first step, the squared linewidth $\Gamma_m^2$ is plotted as a 
function of pump power and fitted linearly with $Y=mX+c$, where $Y$ is the square of the linewidth $\Gamma_m$ measured using the Lorentzian fit, $X$ is the pump + probe power used in the experiment, $m$ and $c$ is the slope and intercept of the linear fit respectively. The square root of the intercept of this linear fit yields 
the total zero-power linewidth $\Gamma_0 =\sqrt{c}$, which contains contributions 
from both homogeneous and inhomogeneous broadening mechanisms. These two 
classes add in quadrature \cite{demtroder2013laser},
\begin{equation}
    \Gamma_0^2 = \Gamma_h^2 + \Gamma_{Ih}^2,
    \label{eq:quadrature}
\end{equation}
where $\Gamma_h$ is the total homogeneous linewidth, comprising the 
natural linewidth $\Gamma = 1.42$~MHz 
\cite{J_Marek_1980, glaser2020absolute, safronova2011critically} and the transit-time broadening 
 $\Gamma_{\rm transit}$ \cite{achar2025direct}. For the beam dimensions $2.985\times1.955$~mm$^2$ 
at 82.02$~\pm~ 0.73^\circ$C, $\Gamma_{\rm transit} \approx 0.44$~MHz, giving 
$\Gamma_h \approx 1.86$~MHz. 
\begin{figure}[h!]
    \centering
    \includegraphics[width=1\linewidth]{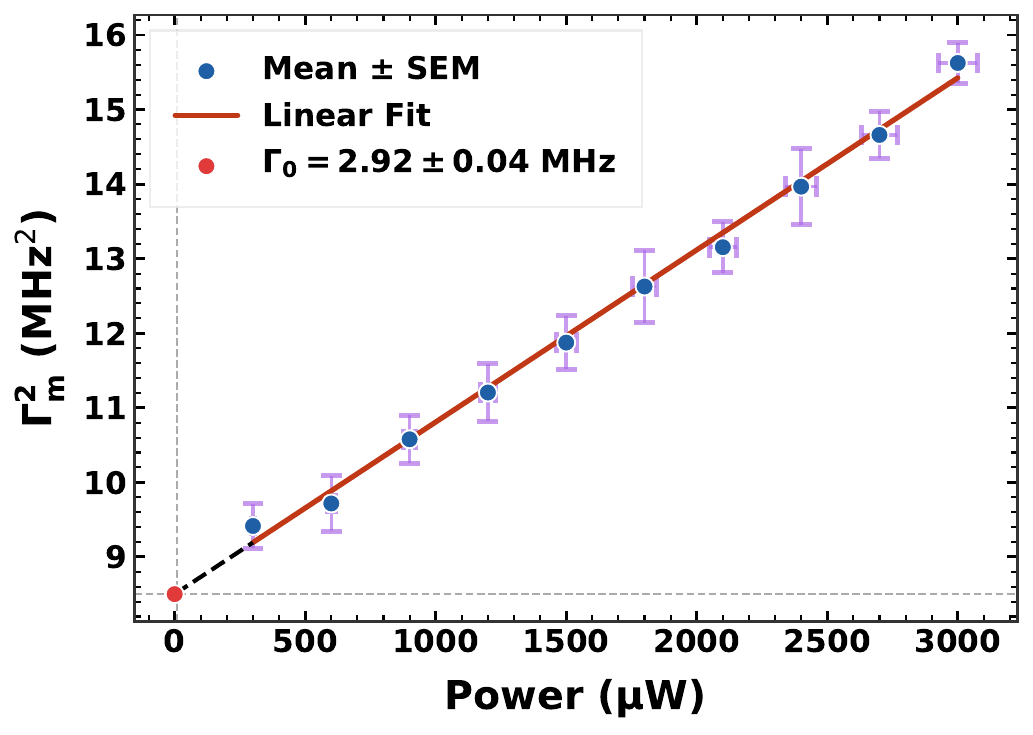 }
    \caption{Measured linewidth square $\Gamma_m^2$ of the $^{85}$Rb 
    $F=3\rightarrow F'=4$ transition as a function of pump power. 
    The solid line is a linear fit, 
    from whose intercept the zero-power linewidth $\Gamma_0$ is 
    extracted. Each data point represents the mean $\pm$ SEM of three 
    independent measurements. The horizontal error bars reflect the 
    calibration uncertainty of the optical power meter 
    (Thorlabs PM400), and the vertical error bars reflect the 
    statistical uncertainty in the fitted linewidth across the repeated 
    datasets.}
    \label{fig:Rb85_intercept}
\end{figure}
The inhomogeneous contribution $\Gamma_{Ih}$, 
arising from residual Doppler background and stray magnetic fields, is 
then determined from the measured $\Gamma_0$ via:
\begin{equation}
    \Gamma_{Ih} = \sqrt{\Gamma_0^2 - \Gamma_h^2}.
    \label{eq:residual}
\end{equation}
The residual inhomogeneous contribution is assumed to remain approximately constant over the investigated power range, whereas the homogeneous component follows the power-broadening dependence. In the second step, the inhomogeneous contribution $\Gamma_{Ih}$ extracted 
from Eq.~(\ref{eq:residual}) is subtracted from each measured linewidth 
data point across the full pump power range, isolating the purely 
homogeneous linewidth at each power. This is physically motivated by the fact that power broadening acts exclusively on the homogeneous (Lorentzian) component of the linewidth \cite{demtroder2013laser}, an applied optical 
field broadens only the coherence decay rate of the transition, leaving the inhomogeneous contribution unaffected. The resulting homogeneous 
linewidth data are then fitted to the power-broadening equation,
\begin{equation}
    \Gamma_h(I) = \Gamma_h\sqrt{1 + \frac{I}{I_{\rm sat}}},
    \label{eq:power_broadening}
\end{equation}
with $I_{\rm sat}$ as the sole free parameter, yielding the saturation 
intensity directly. This two-step procedure ensures that the extracted 
$I_{\rm sat}$ is determined purely from the power-dependent homogeneous 
response of the transition, and is not biased by the inhomogeneous 
broadening present in the raw linewidth data.\\

\begin{figure}[htbp]
    \centering
    \includegraphics[width=1\linewidth]{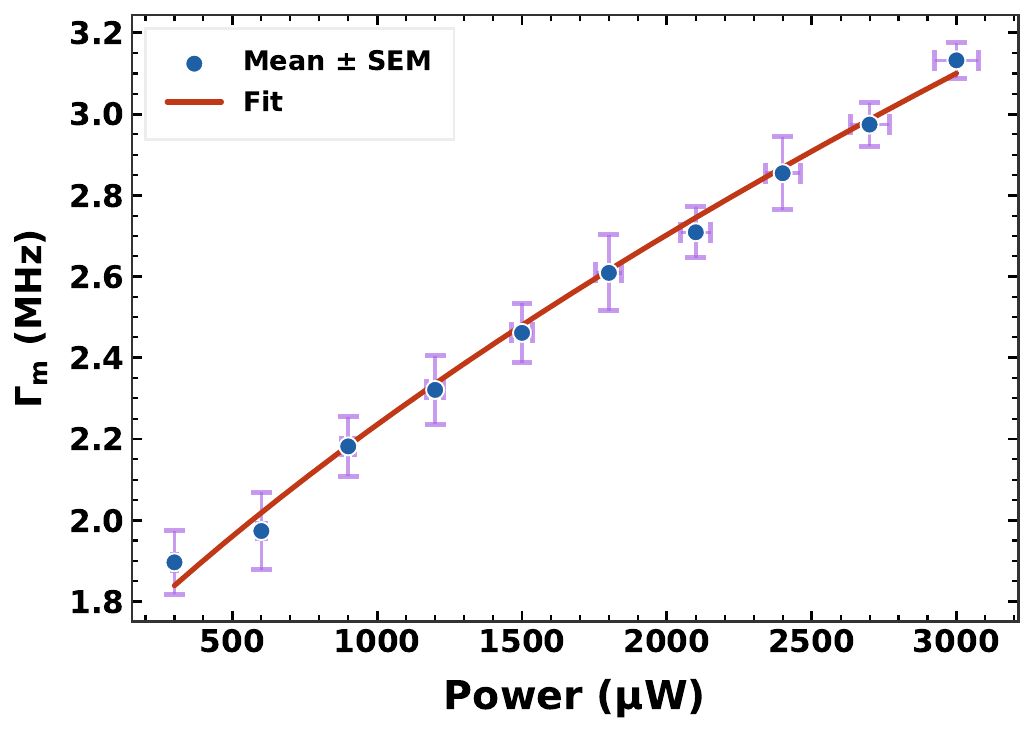}
    \caption{Measured linewidth $\Gamma_m$ of the $^{85}$Rb $F=3\rightarrow F'=4$ saturated absorption resonance as a function of optical power. The solid curve represents the power-broadening fit used to extract the saturation intensity. Each data point represents the Mean $\pm$ SEM of three 
    independent measurements. The horizontal error bars reflect the 
    calibration uncertainty of the optical power meter 
    (Thorlabs PM400), and the vertical error bars reflect the 
    statistical uncertainty in the fitted linewidth across the repeated 
    datasets.}
    \label{fig:Rb85_Isat}
\end{figure}
The squared linewidth $\Gamma_m^2$ as a function of pump power (see Fig.~\ref{fig:Rb85_intercept}) for the $^{85}$Rb $F=3\rightarrow F'=4$ transition measured at 82.02$~\pm~ 0.73^\circ$C using an elliptical beam with dimensions of $2.985 \pm 0.05 \times 1.955 \pm 0.04$ mm$^2$. From the intercept of the linear fit, a zero-power linewidth of $\Gamma_0 = 2.92 \pm 0.04~\mathrm{MHz}$ is obtained, the inhomogeneous contribution ($\Gamma_{Ih}$) is measured using this $\Gamma_0$, which is subsequently 
subtracted from the linewidth data prior to the power-broadening fit.\\

Fig.~\ref{fig:Rb85_Isat} shows the resulting homogeneous linewidth as a function of pump power, after subtraction of $\Gamma_{Ih}$. The solid curve is a fit to Eq.~(\ref{eq:power_broadening}) 
with $I_{\rm sat}$ as the only free parameter. The same measurement procedure was applied to the $^{87}$Rb $F=2 \rightarrow F'=3$ transition.\\


The systematic uncertainty associated with the measurement of the saturation intensity was evaluated by considering the dominant experimental parameters that influence the extracted value. The uncertainty budget includes contributions from the frequency calibration, beam diameter measurement, and optical power measurement.\\
\begin{table*}[t]
\caption{Summary of the systematic uncertainty budget for the measured saturation intensity of the $^{85}$Rb and $^{87}$Rb transitions. The contributions from frequency calibration, beam diameter measurement, and optical power calibration are evaluated.}
\label{tab:uncertainty}
\centering
\renewcommand{\arraystretch}{1.3}
\setlength{\tabcolsep}{8pt}

\begin{tabular}{lccccc}
\hline\hline
\textbf{Parameter} & \textbf{Isotope}&
\textbf{Nominal Value} &
\textbf{Uncertainty} &
\boldmath$\sigma_N$ (\textbf{m}$^{-3}$) &
\textbf{Relative}\\
& & & & &  \textbf{ Uncertainty (\%)}\\
\hline
Frequency Calibration & $^{85}$Rb &
1.5 GHz &
1 MHz &
0.61&
2.40 \\

Beam diameter & $^{85}$Rb &
2.416 mm &
0.045 mm &
0.85 &
3.33 \\

Power & $^{85}$Rb &
300-3000 $\mu$W &
5\% &
0.45 &
1.78 \\
Frequency Calibration & $^{87}$Rb &
1.5 GHz &
1 MHz &
0.57 &
2.42 \\

Beam Diameter & $^{87}$Rb &
2.416 mm &
0.045 mm &
0.73 &
3.02 \\

Power & $^{87}$Rb &
300-3000 $\mu$W &
5\% &
0.44 &
1.84 \\
\hline
\textbf{Total systematic uncertainty} &
&
&
$\sigma_{\rm sys}$ ($^{85}$Rb) &
\textbf{4.47} \\
&
&
&
$\sigma_{\rm sys}$ ($^{87}$Rb) &
\textbf{4.28} \\
\hline\hline
\end{tabular}
\end{table*}
The uncertainty arising from the frequency calibration was estimated by perturbing the frequency scale by its calibration uncertainty ($\pm 1~\mathrm{MHz}$) and evaluating the corresponding variation in the extracted saturation intensity using a central finite-difference approach. Similarly, the uncertainty due to the beam diameter was determined by varying the measured beam diameter within its experimental uncertainty and repeating the complete fitting procedure.\\

The systematic uncertainty associated with the optical power measurement was evaluated using the specified calibration uncertainty of the power meter (2.5\%). A numerical sensitivity analysis was performed by perturbing the measured pump powers within the calibration uncertainty and repeating the entire fitting procedure to determine the corresponding change in the extracted saturation intensity. The resulting variation was taken as the systematic uncertainty associated with the optical power measurement.\\

No additional systematic uncertainty was assigned to the baseline correction. The saturation intensity was extracted from Doppler-free saturated absorption spectra, for which the detector dark current contributes only as a constant offset. Since this constant background is removed during data acquisition and baseline subtraction, it does not alter the linewidth or the extracted saturation intensity. Consequently, the systematic uncertainty arising from baseline correction was considered negligible and set to zero. Furthermore, the AC Stark (light) shift was not included in the uncertainty budget since, for the optical powers used in this work, the estimated shift is approximately $200$ Hz, which is negligible compared with both the natural linewidth and the experimental frequency calibration uncertainty.\\

Assuming that the individual systematic contributions are independent, the total systematic uncertainty was obtained by combining the individual components in quadrature, $\sqrt{
\sigma_{\mathrm{freq}}^{2}
+
\sigma_{\mathrm{beam}}^{2}
+
\sigma_{\mathrm{power}}^{2}
}$. The resulting uncertainty budget is summarized in Table \ref{tab:uncertainty}. The total relative systematic uncertainties were found to be (4.47\%) for $^{85}$Rb and (4.28\%) for $^{87}$Rb.\\

The consistency of the measured saturation intensity across a wide range of 
cell temperatures (49.35$~\pm~0.25$--82.02$~\pm~0.73^\circ$C) confirm that the measurements are free from systematic offsets due to beam geometry or vapor density. A comparison of the measured and theoretical saturation intensities for both isotopes is summarized in Table~\ref{tab:isat}.
\begin{table}[h!]
\centering
\caption{Comparison of the theoretical and experimentally measured saturation intensities for the $^{87}$Rb and $^{85}$Rb transitions. The experimental values represent the mean saturation intensity obtained from measurements at four different vapor temperatures (49.35$\pm$0.25$^\circ$C, 59.03$\pm$0.37$^\circ$C, 72.90$\pm$0.57$^\circ$C, and 82.02$\pm$0.73$^\circ$C). The reported uncertainties include both the statistical uncertainty of the mean (SEM) and the combined systematic uncertainty.}
\label{tab:isat}
\setlength{\tabcolsep}{8pt}
\begin{tabular}{lcc}
\toprule
Transition &
Theory & Experiment \\
& (mW/cm$^2$) & (mW/cm$^2$) \\
\midrule
$^{87}$Rb ($F=2\rightarrow F'=3$) &
23.45 &
$23.51\pm1.03$ \\
$^{85}$Rb ($F=3\rightarrow F'=4$) &
25.54 &
$25.39\pm1.16$ \\
\bottomrule
\end{tabular}
\end{table}
It is worth noting that the saturation intensities at 420~nm are 
approximately 6--7 times larger than those of the well-studied $D_2$ line at 
780~nm ($I_{\rm sat} \approx 3.58$~mW/cm$^2$ for $^{87}$Rb, 
$F=2 \rightarrow F'=3$). This increase arises primarily from the weaker 
effective coupling strength of the $5S_{1/2} \rightarrow 6P_{3/2}$ 
transition due to the cubic dependence of the spontaneous emission rate on frequency and the significantly reduced branching ratio back to the ground state (23\% compared to nearly 100\% for the $D_2$ line). This implies that 
significantly higher optical power is required to saturate the 420~nm  transition, which has practical implications for 
applications in compact atomic sensors.

\subsection{Temperature Dependence of the Linewidth and Signal Amplitude}
\begin{figure}[htbp]
    \centering
    \includegraphics[width=0.9\linewidth]{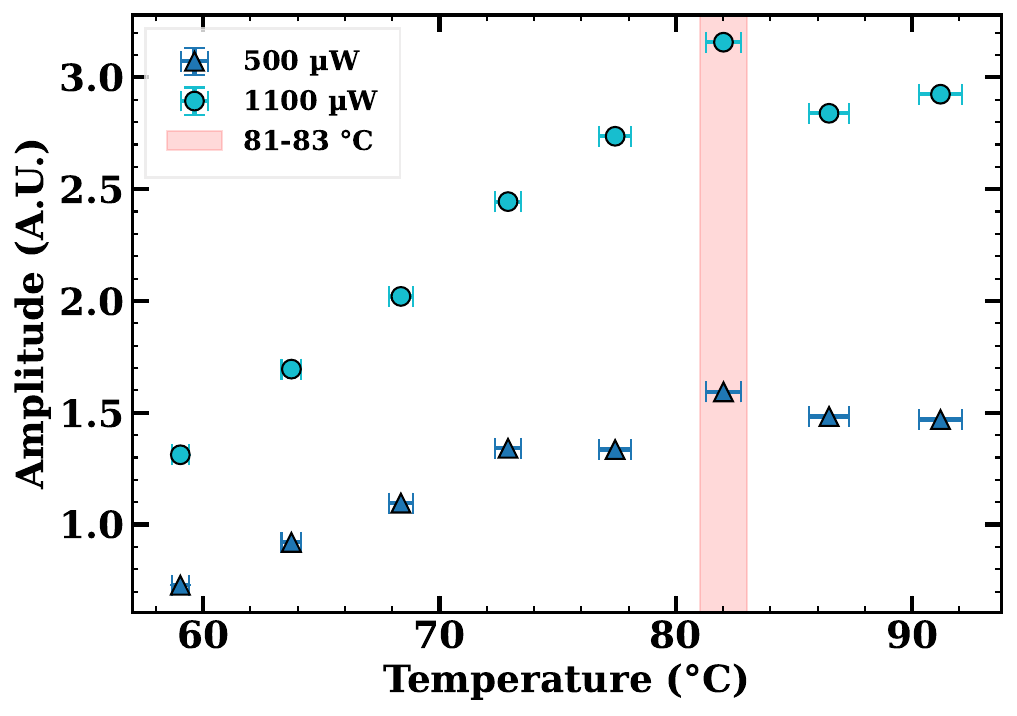      }
    \caption{The amplitude variation of the Lamb-dip $^{85}$Rb,$~F=3\to F' = 4$ Lamb-dip resonance as a function of the measured vapor-cell temperature for the 100-mm-long vapor cell for pump powers of 500~$\mu$W and 1100~$\mu$W. The x-axis values correspond to the mean temperature measured by six calibrated sensors attached to the vapor cell, with horizontal error bars indicating the SEM. Vertical error bars represent the SEM of five independent amplitude measurements at each temperature. The shaded region marks the optimal operating temperature range yielding the highest SNR.}
    \label{fig:amplitude_temp}
\end{figure}

To characterize the optimal operating conditions for the vapor cell 
at 420~nm, the SAS Lamb-dip linewidth and amplitude were measured as a 
function of cell temperature over the range 59.03$~\pm~0.37$ - 91.20$~\pm~0.90^\circ$C for 
$\sim$50~$\mu$W probe measurements, at fixed pump powers of $\sim$500, and 
$\sim$1100~$\mu$W. Fig.~\ref{fig:amplitude_temp} shows the Lamb-dip amplitude of the $F=3\to F'=4$ transition in $^{85}Rb$ as a function of temperature. As expected, the amplitude increases monotonically with temperature for all pump powers, consistent with the exponential increase in Rb vapor pressure and hence atomic number density with temperature. Higher pump powers produce proportionally larger Lamb-dip amplitudes due to greater population saturation.
At further higher temperatures after ($\gtrsim$82.02$ ~\pm~ 0.73^\circ$C), increased atomic collisions and residual pressure broadening from Rb-Rb collisions force the amplitude to reduce.
\begin{figure}[htbp]
    \centering
    \includegraphics[width=0.9\linewidth]{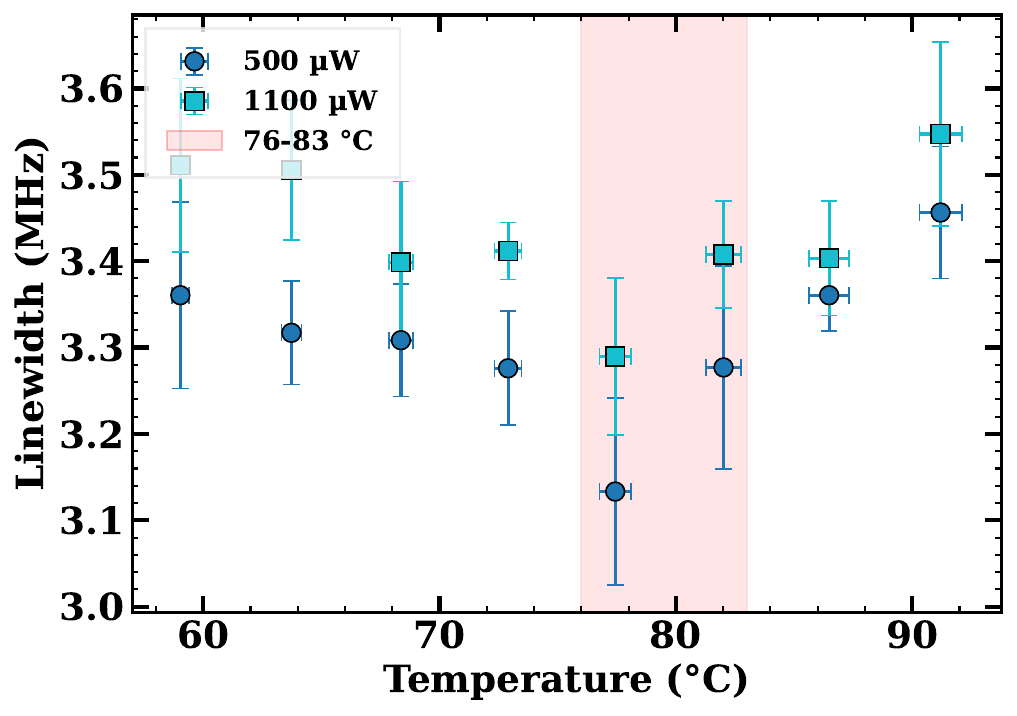}
    \caption{Temperature dependence of the Lamb-dip linewidth of the $^{85}$Rb $5S_{1/2}(F=3)\rightarrow6P_{3/2}(F'=4)$ transition measured in a 100-mm-long vapor cell for pump powers of 500~$\mu$W and 1100~$\mu$W. The x-axis values correspond to the average vapor-cell temperature measured by six calibrated temperature sensors mounted at different locations on the cell, with horizontal error bars representing the standard error of the mean (SEM). The vertical error bars denote the SEM of the linewidth obtained from five statistically independent measurements at each temperature. The shaded region (76--83$^\circ$C) indicates the optimal operating temperature range, where the minimum linewidth is observed. }
    \label{fig:linewidht_temp}
\end{figure}
Fig.~\ref{fig:linewidht_temp} shows the corresponding Lamb-dip linewidth of the $F=3\to F'=4$ transition in $^{85}Rb$ as a function of temperature. In contrast to the amplitude, the linewidth exhibits a non-monotonic temperature dependence with a broad minimum in the range of 76-83$^\circ$C. At lower temperatures, the linewidth is dominated by the power broadening relative to the small Lamb-dip signal. 
As the temperature increases, the increasing atomic number density enhances the signal amplitude and the effective SNR and reduces the apparent fitted linewidth toward its natural minimum. At higher temperatures around ($\gtrsim$82.02$~\pm~ 0.73^\circ$C), increased atomic collisions and residual pressure broadening from Rb-Rb collisions cause the linewidth to rise again.\\

From these measurements, the optimal operating temperature for the Rb vapor cell at 420~nm is identified as near 82.02$~\pm~ 0.73^\circ$C, where the Lamb-dip linewidth is minimized while the signal amplitude remains sufficiently large for reliable spectroscopic measurements. This optimal operating point is consistent across both $^{85}$Rb and $^{87}$Rb isotopes and across both probe power conditions ($\sim$20 and $\sim$50~$\mu$W), indicating the intrinsic property of the Rb vapor rather than a measurement artifact.\\

\subsection{Hyperfine Structure Constants of the $6P_{3/2}$ State}
The hyperfine structure constants $A$ and $B$ of the $6P_{3/2}$ state were determined by measuring the frequency intervals between the resolved Lamb-dip features in the saturated absorption spectra.\\
\begin{figure}[h]
    \centering
    \includegraphics[width=1\linewidth]{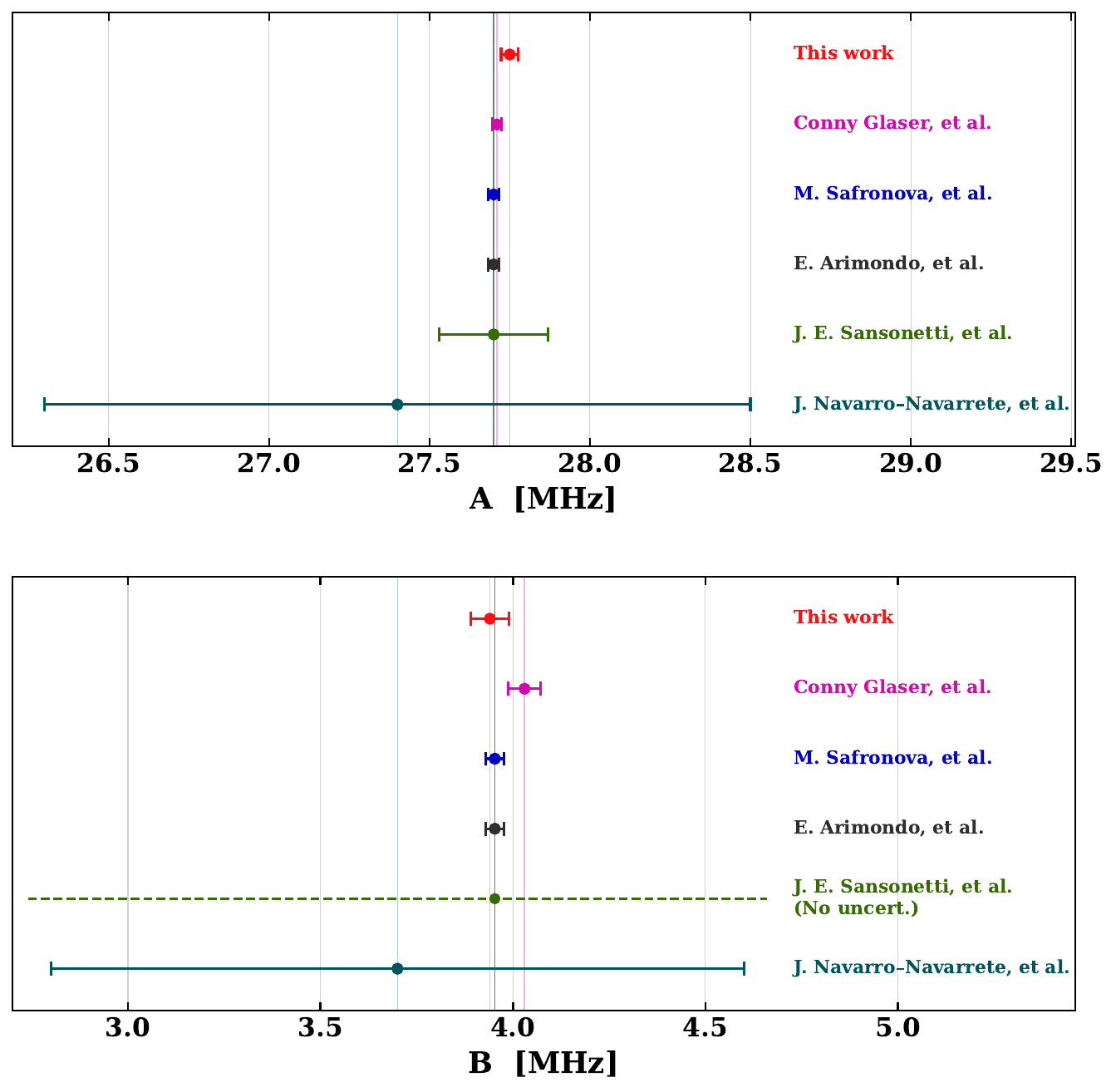}
    \caption{Measured magnetic-dipole ($A$) and electric-quadrupole ($B$) hyperfine constants of the $6P_{3/2}$ state in $^{87}$Rb. The results were obtained from four independent measurements, each consisting of 50 statistically independent datasets recorded at 82.02$~\pm~ 0.73^\circ$C, with probe power of $\sim$50 $\mu$W and pump power at $\sim$1100 $\mu$W, laser beam diameter of $2.985~\pm~ 0.05$ mm and $1.955~\pm~ 0.04$ mm for a 100 mm vapor cell. Error bars denote the standard error of the mean (SEM).}
    \label{fig:Hyperfine_Rb_87_B}
\end{figure}

 The frequency axis was calibrated using the transmission fringes of an FPI with a known free spectral range recorded simultaneously with the SAS signal. To evaluate the statistical reproducibility of the measurement, 50 statistically independent spectra were recorded at vapor-cell temperature 82.02$~\pm~ 0.73^\circ$C using a probe power of approximately 50~$\mu$W and a pump power of approximately 1100~$\mu$W. The hyperfine constants were extracted independently from each spectrum by fitting the measured transition frequencies. The final values correspond to the arithmetic mean of the 50 independent measurements, while the quoted uncertainties represent the standard error of the mean (SEM, $1\sigma$).\\
 
Among the investigated temperatures, the spectra recorded at 82.02$~\pm~ 0.73^\circ$C exhibited the narrowest linewidths and the highest SNR. Consequently, the hyperfine constants obtained at this temperature are quoted as the final experimental values, while measurements at the other temperatures yielded statistically consistent results within the quoted uncertainties.\\
\begin{figure}[H]
    \centering
    \includegraphics[width=1\linewidth]{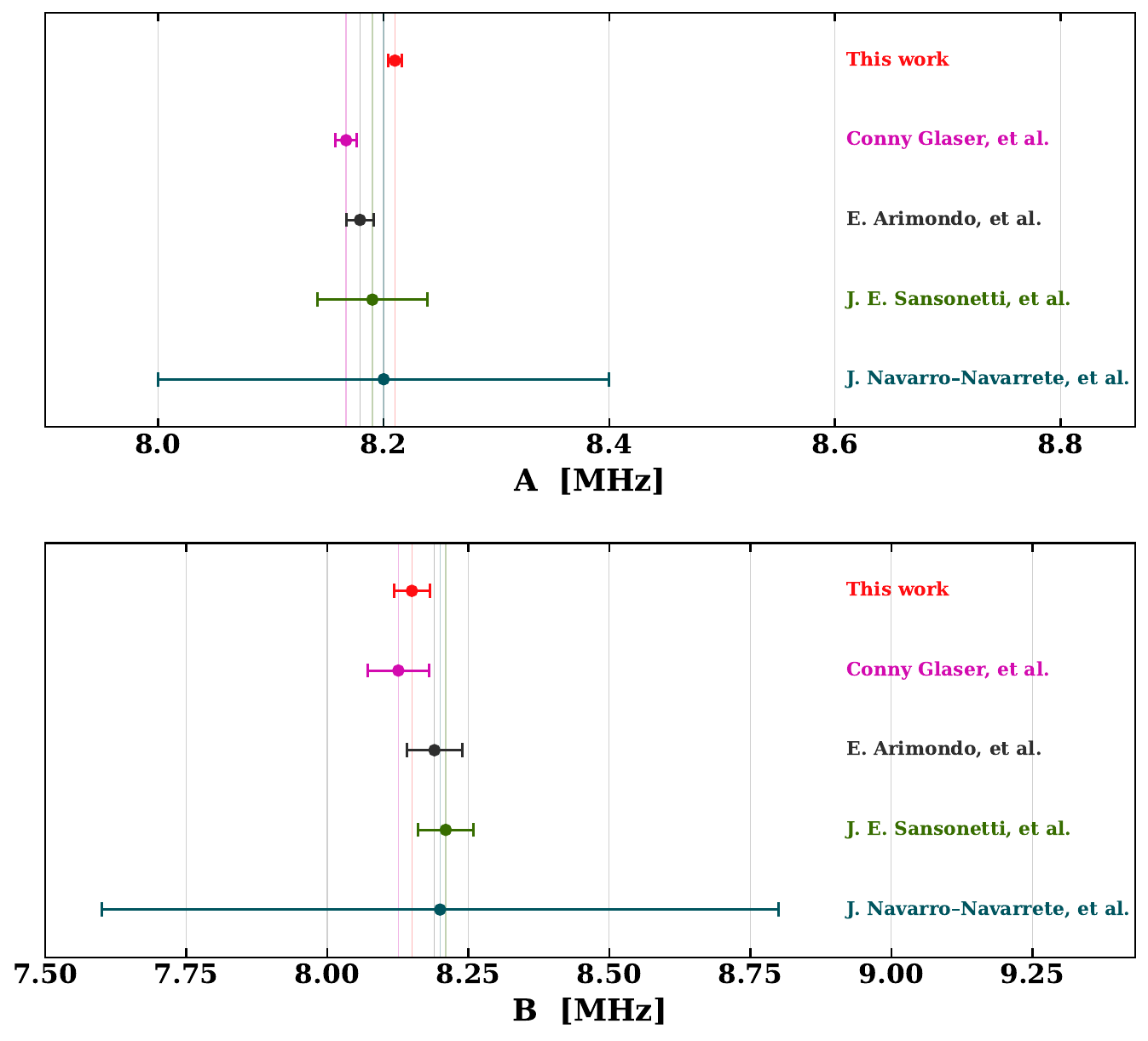}
    \caption{Measured magnetic-dipole ($A$) and electric-quadrupole ($B$) hyperfine constants of the $6P_{3/2}$ state in $^{85}$Rb. The results were obtained from four independent measurements, each consisting of 50 statistically independent datasets recorded at 82.02$~\pm~ 0.73^\circ$C, with probe power $\sim$50 $\mu$W, pump power $\sim$1100 $\mu$W, and a laser beam diameter of $2.985 ~\pm~ 0.05$ mm and $1.955 ~\pm~ 0.04$ mm for a 100 mm vapor cell. Error bars denote the standard error of the mean (SEM).}
    \label{fig:Rb85_hyperfine}
\end{figure}
Fig.~\ref{fig:Hyperfine_Rb_87_B} and Fig.~\ref{fig:Rb85_hyperfine} summarizes the measured hyperfine constants of the $6P_{3/2}$ state of $^{87}$Rb and $^{85}$Rb respectively. The measured A and B coefficients for both isotopes are summarized alongside literature values in Table~\ref{tab:hyperfine}.
\begin{widetext}
\begin{center}
\renewcommand{\arraystretch}{1.2}

\begin{table}[H]
\centering
\caption{Measured hyperfine structure constants $A$ and $B$ of the
$6P_{3/2}$ state of $^{85}$Rb and $^{87}$Rb compared with selected
literature values. Numbers in parentheses denote the one-standard-deviation ($1\sigma$) uncertainty in the last quoted digits. All values are given in MHz.}
\label{tab:hyperfine}

\setlength{\tabcolsep}{20pt}
\begin{tabular}{|c|c|c|c|c|}
\hline
\multirow{2}{*}{\textbf{Source}} &
\multicolumn{2}{c|}{\textbf{$^{87}$Rb}} &
\multicolumn{2}{c|}{\textbf{$^{85}$Rb}} \\
\cline{2-5}
& \textbf{$A$} & \textbf{$B$} & \textbf{$A$} & \textbf{$B$} \\
\hline

This work &
$\mathbf{27.75(03)}$ &
$\mathbf{3.94(05)}$ &
$\mathbf{8.21(006)}$ &
$\mathbf{8.15(03)}$ \\

Conny Glaser, \textit{et al.} \cite{glaser2020absolute} &
27.71(15) & 4.03(04) & 8.16(01) & 8.13(05) \\
E. Arimondo, \textit{et al.} \cite{arimondo1977experimental} &
27.7(02) & 3.95(02) & 8.18(01) & 8.19(04) \\
M. Safronova, \textit{et al.} \cite{safronova2011critically} &
27.700(017) & 3.953(024) & - & - \\
J.E. Sansonetti, \textit{et al.} \cite{sansonetti2006wavelengths} &
27.700(17) & 3.953(24) & 8.179(12) & 8.190(49) \\
J. Navarro-Navarrete, \textit{et al.} \cite{navarro2019doppler} &
27.4(110) & 3.7(90) & 8.2(40) & 8.2(60) \\
\hline
\end{tabular}
\end{table}
\end{center}
\end{widetext}
\section{Conclusion} \label{sec:conclusion}
We have presented a systematic experimental investigation of the 
$5S_{1/2}\rightarrow6P_{3/2}$ transition in Rb at 420~nm using 
Doppler-free saturated absorption spectroscopy, addressing the saturation intensity, temperature dependence of the Lamb-dip signal, and the hyperfine structure constants of the $6P_{3/2}$ state.\\

The saturation intensity was measured for both isotopes using  a linear fit to $\Gamma^2_m$ versus power and a nonlinear power-broadening fit to the Lamb dip 
FWHM across four cell temperatures and two beam configurations. The weighted mean values, 
$I_{\rm sat}(^{87}{\rm Rb}) = 23.54\pm1.03$~mW/cm$^2$ and 
$I_{\rm sat}(^{85}{\rm Rb}) = 25.39\pm1.16$~mW/cm$^2$, are in good 
agreement with the theoretical predictions of 23.45 and 25.54~mW/cm$^2$, respectively. These values are approximately 6-7 times larger than those of the $D_2$ line at 780~nm, which reflects the weaker effective coupling strength due to the 23\% branching ratio of the $6P_{3/2}$ state back to the ground state.\\

The temperature characterization identifies 81-83$^\circ$C as the optimal operating point, where the signal amplitude is sufficiently large, and the Lamb-dip linewidth is minimized for reliable spectroscopic measurements. The hyperfine constants measured from the resolved Lamb-dip frequency intervals, $A~(^{87}{\rm Rb}) = 27.75\pm0.03$~MHz, $B~(^{87}{\rm Rb}) = 3.94\pm0.05$~MHz, $A~(^{85}{\rm Rb}) = 8.21\pm0.006$~MHz, and $B~(^{85}{\rm Rb}) = 8.15\pm0.03$~MHz, are in good agreement with previously reported values \cite{sansonetti2006wavelengths, navarro2019doppler, glaser2020absolute, safronova2011critically}.\\

These saturation intensities and optimal operational parameters constitute quantitative standards for further experimentation on the Rb transition at 420~nm, including laser stabilization at blue light, two-photon spectroscopy, Rydberg excitation mechanisms, and miniature warm-vapor atomic detectors.

\subsection*{Acknowledgments}
Shivam Sinha gratefully acknowledges the financial assistance provided by IIT Tirupati, India, facilitated through the half-time research assistantship (HTRA). Sumit Achar gratefully acknowledges financial support from the Council of Scientific \& Industrial Research (CSIR, Govt. of India) through a Senior Research Fellowship (SRF). Arijit Sharma acknowledges financial support from IIT Tirupati through the CAMOST grant, ANRF (erstwhile SERB) SRG Grant SRG/2020/001049, and DST NQM project grant DST/QTC/NQM/QComm/2024/2(C).\\

\subsection*{Conflict of interest}
The authors declare that they have no conflict of interest.

\subsection*{Data availability}
The data that support the findings of this study are available upon a reasonable request from the authors.

\bibliography{reference}
\end{document}